\documentclass[,final]            
  {aipproc}
\layoutstyle{8x11double}

\begin{document}

\title{High-frequency QPOs as a problem in physics: 
		  non-linear resonance}

\author{W{\l}odek Klu{\'z}niak}{
  address={Institute of Astronomy,
   Zielona G\'ora University, ul. Lubuska 2, 65-265 Zielona G\'ora, Poland},
  altaddress={Copernicus Astronomical Center,
   ul. Bartycka 18, 00-716 Warszawa, Poland},
}
\author{Marek A. Abramowicz}{
  address={Department of Astrophysics, Chalmers University, G\"oteborg, Sweden}
}
\author{William H. Lee}{
  address={Instituto de Astronom\'{\i}a, 
Universidad Nacional Aut\'{o}noma de M\'{e}xico, \\ 
Apdo. Postal 70-264, Cd. Universitaria, 
M\'{e}xico D.F. 04510}
}
\begin{abstract}
The presence of a kHz frequency in LMXBs has been expected
from scaling laws, by analogy with the QPO phenomenon in HMXB X-ray pulsars.
Interpretation of the two kHz frequencies, observed in accreting
neutron stars,  in terms of non-linear resonance
in strong-field gravity led to the prediction
of twin  QPOs in black hole systems, in a definite frequency ratio
(such as 2/3).
The imprint of a subharmonic of the 401 Hz rotation rate
in the frequencies of the QPOs detected in
the accreting millisecond pulsar is at once a signature of non-linear
resonance and of coupling between accretion disk modes
and the neutron star spin.

\end{abstract}

\maketitle

%%%%%%%%%%%%%%%%%%%%%%%%%%%%%%%%%%%%%%%%%%%%
%% Paper
%%%%%%%%%%%%%%%%%%%%%%%%%%%%%%%%%%%%%%%%%%%%

\section{Introduction}

This is a story of the origin of high-frequency quasi-periodic oscillations
(HF QPOs) observed in the X-ray flux of low-mass X-ray binaries (LMXBs).
We believe that they really are oscillations, resonant oscillations of
a body of fluid nearly in equilibrium---the accretion disk.
Accretion disk are so poorly understood that their detailed
numerical and analytic modeling has failed to reveal phenomena
approximating the behavior of real HF QPOs.
Even so, they must exhibit behavior generally observed in any sufficiently
complicated system, e.g., resonances. 

Physics has principles and techniques allowing
a meaningful discussion of the behavior of systems whose detailed structure
is not known. The application of these principles has allowed us
to successfully predict \cite{kmw,ka00} millisecond variability in LMXBs,
at a time
when at most $\sim 50\,$Hz oscillations had been seen, and the appearance
of pairs of hHz QPOs with rational (as in 3:2) frequency ratios in black holes,
at a time when only single QPOs had been reported.
We see no reason to abandon these principles now, when 
the field of QPO modeling  has been thrown into turmoil with the discovery
\cite{rudy} of
a subharmonic frequency difference in the accreting millisecond pulsar.
This is where they take us. High-frequency QPOs are an accretion
disk phenomenon reflecting fundamental properties of gravity.
A spinning neutron star may directly perturb the accretion disk
and influence the frequencies of its
oscillation via a non-linear coupling.

\section{An accretion-disk phenomenon}
There is little doubt that QPOs reflect rapid variations of flow in
accretion disks. The argument is simple. Essentially the same phenomenon
is seen in three types of mass-transfering binary systems.
In all cases the donor is a low-mass star and the bright source
is a compact object in state of accretion: a white dwarf, a neutron
star, or a black hole. The observed time-scale 
of variation depends on which of the three,
so the donor canot be responsible for (and is not expected to exhibit)
the rapid variability. The white dwarfs have a surface capable of emission
and an accretion disk around it. The neutron stars likewise, and also
an occasional jet. Black holes (microquasars) have an accretion disk and a jet
but no surface. Therefore, if the phenomenon has the same origin
(see e.g., \cite{warner} for a comparative phenomenology) it can only
occur in the single structure that the three types of systems have in common:
the accretion disk.

\section{Fundamental or accidental?}
Some fifteen years ago, when quasiperiodic variability of luminosity had
already been detected in white dwarfs (cataclysmic variables),
in strongly magnetized accreting neutron stars (X-ray pulsars)
and in weakly (if at all) magnetized accreting neutron star,
the (tacit) question of the day was whether these were separate phenomena
reflecting accidental properties of the systems, such as their rotation
rates, magnetic fields, and the such, or whether instead, this was a
phenomenon related to the fundamental frequency of the system,
$(2\pi)\nu_K=\sqrt{GM/r^3}$ \cite{bath}.
Much confusion arose, because the highest frequency then observed in LMXBs 
was only $\sim50\,$Hz$\,<<\nu_K$. Ignoring this particular QPO,
the authors of ref. \cite{kmw} opted for the Keplerian model of
the sub-Hertz QPO reported in the X-ray pulsar EXO 2030 + 375
\cite{ang}, and suggested,
by analogy, that in a neutron star with magnetic field $<10^8\,$G
a frequency close to the orbital frequency in the marginally stable orbit of
general relativity, $\nu_K(r_{\rm ms})\approx2.2{\,\rm kHz\,}(M_\odot/M)$,
should be observed.
Of course, a frequency of about 2 kHz per inverse solar mass is
precisely what has been discovered with RXTE in LMXBs, so it is reasonable
to examine in detail the line of reasoning. 

\section{Scaling with radius}
The first assumption \cite{kmw} was
that the frequency should (more or less) exhibit the fundamental
scaling of Kepler's third law, $\nu\propto r^{-3/2}$.
This scaling is roughly followed.
Indeed, for X-ray pulsars, where the disk ends close to the magnetospheric
radius $r\approx10^8\,$cm, the frequency is $\nu\sim0.2$ Hz \cite{ang}
(see also Mark Finger's talk in this conference \cite{finger}), 
in white dwarfs with radius $r\sim10^9\,$cm the frequencies cluster at
$\nu\sim0.04\,$cm, and for neutron stars in LMXBs, with their stellar
radius $\sim r_{\rm ms}\sim10^6\,$cm \cite{kw}, the QPO frequency is
$\nu\sim1\,$ kHz. Note that in the first two cases the radius depends on the
stellar properties. For the LMXB neutron stars, the inner disk radius
was expected to be determined by properties of Einstein's gravity
\cite{kw,sush,kmw}.
 
\section{Scaling with mass}
The second assumption was that once the direct influence of the stellar
surface and magnetic field is negligible, the characteristic frequency
is determined by gravity, so that the characteristic radius determining the
QPO properties is proportional to $GM/c^2$, and hence
$\nu_{\rm QPO}\propto 1/M$,
the inverse mass of the neutron star (or a  black hole). For an elaboration
of this point see the companion contribution by Abramowicz and Klu\'zniak
\cite{paboston}.

This scaling is satisfied quite well. For $\sim 1M_\odot$ neutron stars,
$\nu_{\rm QPO}\approx 1\,$kHz, for a $\sim 10M_\odot$ black hole
$\nu_{\rm QPO}\approx0.1\,$kHz, and for the $ 4\cdot10^6M_\odot$ black hole
at the Galactic center the recently reported period (of an infrared flare)
 is 17 minutes \cite{genzel}, i.e.,
 $\nu_{\rm QPO}\approx1\,$mHz.
It seems quite safe to assume that the HF QPOs are closely related to
orbital frequencies in Einstein's gravity.

\section{Two variable characteristic frequencies}
The puzzle of twin kHz QPOs in neutron stars is this \cite{ka00}:
the two frequencies have characteristic values, and yet they vary in time
(within minutes).
It is easy enough to find two characteristic frequencies
for neutron stars or black holes, e.g., the maximum radial epicyclic
frequency (Figs. 2,~3 in ref.  \cite{paboston}, this volume),
and  $\nu_K(r_{\rm ms})$, but these frequencies are fixed on human time-scales
(the mass and angular momentum of the neutron star change only
over millions of years).

The resolution of the puzzle, suggested in ref. \cite{ka00,ka01},
is non-linear resonance. Resonance occurs at characteristic frequencies,
for non-linear systems these can vary \cite{lali,mook}.

%======================================================================
%==
%==        Figure 1 OK
%==
%======================================================================

\begin{figure} [!ht]
  \includegraphics[angle=-90,width=78mm]{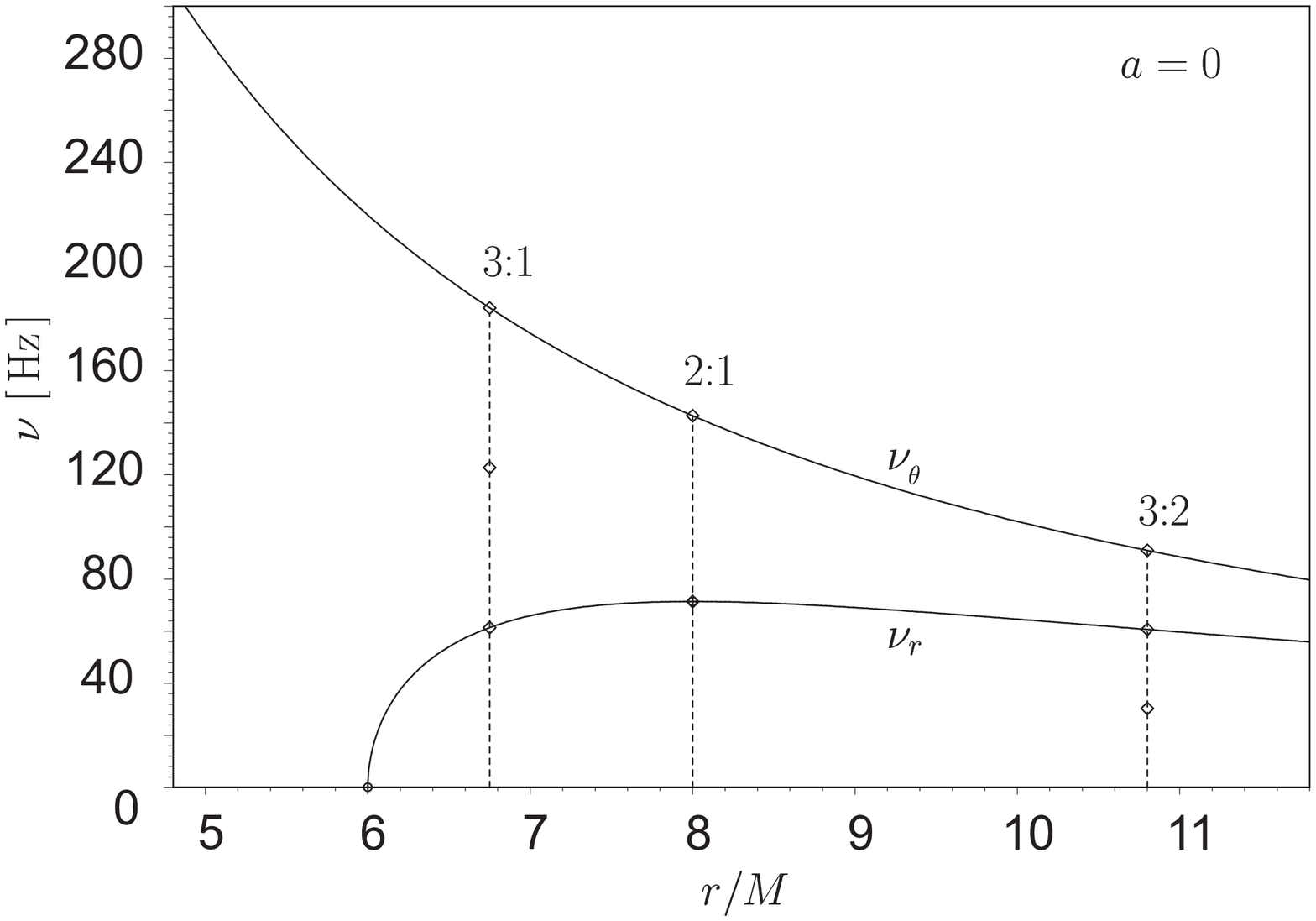}
  \caption{Radial
  $\nu_{\rm r}$, and vertical $\nu_{\theta}$ epicyclic frequencies for a
  non-rotating black hole (spin $a=0$) of mass
  $M  =10\,M_{\odot}$. Location of various possible resonances
  between the two is shown as a function of the dimensionless radius.}
\end{figure}

\section{Epicyclic resonance}
Of course, one still needs to identify the resonant modes.
An intriguing possibility is that the resonance occurs between  
accretion-disk motions occurring at the two epicyclic frequencies.
An even more specific suggestion is that the resonance is parametric
 \cite{ka02,jpn}, which can
occur when the two frequencies are in a $2/n$ ratio, with
$n$ integer, \cite{lali}.
Since the meridional epicyclic frequency is larger than the radial
epicyclic frequency, $\nu_r/\nu_\theta<1$, (Fig.~1), the lowest value
of the integer at which parametric resonance can occur is $n=3$,
assuming that it is driven by small-amplitude radial oscillations.
This would give $\nu_r/\nu_\theta=2/3$ for the ratio of the two frequencies.
A related $g$-mode calculation \cite{shoji} yields $1/\sqrt2$.

\section{Sco X-1}
The published data on Sco X-1 \cite{sco} exhibits clustering of the kHz QPO
frequency ratio near 2/3 \cite{abbk}, suggesting that the twin QPOs
may be related to parametric resonance
between the epicyclic frequencies. However, the correlation line of
the two frequencies has a different slope from 2/3. This can be explained
if the system is slightly off-resonance \cite{akklr}.

%======================================================================
%==
%==        Figure 2 OK
%==
%======================================================================

\begin{figure} [!ht]
  \includegraphics[angle=0,width=78mm]{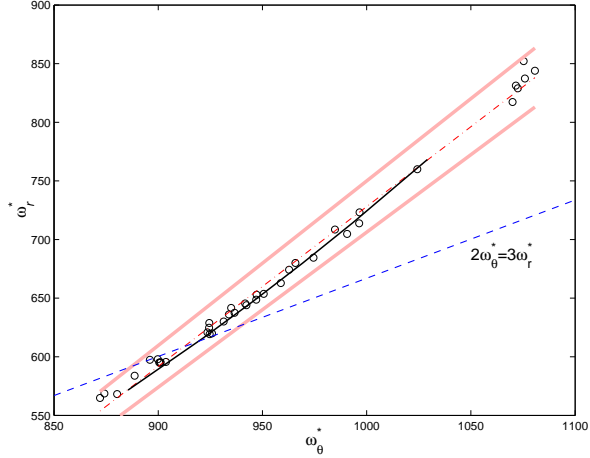}
  \caption{The correlation between the two kHz QPO frequencies
in Sco X-1 (after ref. \cite{akklr}).
 The data (dots) is from van der Klis et al. \cite{sco}.
The continuous line going through the main cluster of points is 
a theoretical calculation of the frequencies expected
slightly off-resonance in a model of parametric resonance
between epicyclic motions in an accretion disk.
}
\end{figure}

\section{Signatures of non-linearity}
A rational, but not a unit, ratio of resonant frequencies,
is expected in non-linear resonance, and the presence of QPO frequencies
in a 2:3 ratio in microquasars
\cite{mcr} is a strong argument in favor of this interpretation
of QPOs. However, simple flute overtones can also give this ratio
\cite{luciano}, see also \cite{ramesh}. 
 The presence of subharmonic frequencies is
a hallmark of  non-linear resonance\cite{mook}.
 Possible subharmonics have already been
reported \cite{rem} in one of the microquasar QPOs, XTE J1550-564,
and their significance duly noted \cite{jpn}.
Strong evidence of non-linearity is afforded by the subharmonic
(in relation to the neutron-star spin) separation of QPO peaks of about 200 Hz
in the 401 Hz accreting pulsar \cite{rudy}.

%======================================================================
%==
%==        Figure 3 OK
%==
%======================================================================
\begin{figure} [!ht]
  \includegraphics[angle=0,
  width=78mm]{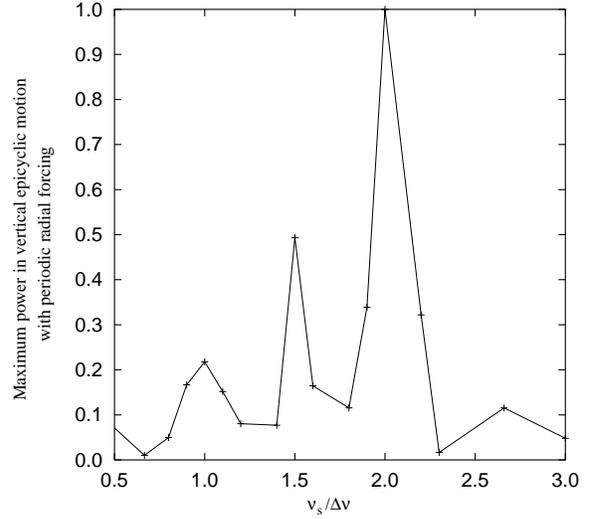}
  \caption{Peak power in the induced vertical oscillations of a
  slender torus perturbed radially at frequency $\nu_s$, as
  a function of the ratio $\nu_{s}/\Delta \nu$, where $\Delta \nu$ is
  the difference between the vertical and radial epicylic frequencies
  at the center of the torus. Three peaks are clearly visible, at 1:1,
  3:2 and 2:1 ratios.}
\label{fftpower}
\end{figure}
\section{Forced resonance}

A spinning neutron star at the center of an accretion disk
is a source of possible perturbations for the latter. These
could be produced either through coupling via the pulsar magnetic
field, or some disturbance on the surface of the star. In each case,
the perturbation would be periodic, repeating at the pulsar spin frequency,
$\nu_s$. Given this fact, one can ask which modes will be
excited in the disk. Dynamical (SPH) calculations of the oscillations of
slender torii, slightly perturbed away from hydrostatic equilibrium,
show that the response to such perturbations occurs primarily at both, the
local radial, and vertical, epicylic frequencies, even if the applied
perturbation is purely radial (or vertical) \cite{pieciu}. 
This demonstrates that a
certain amount of coupling, purely due to pressure, is present. 
We find \cite{lee}, that the magnitude of the response is
dependent on the relation between the spin period and the difference
between the epicyclic frequencies, $\Delta \nu$. Peak response
in vertical oscillations occurs when the difference between the two
epicyclic frequencies is equal to one-half the perturbing
frequency (Figure~\ref{fftpower}). This, $\nu_s/2$, is the frequency
separation observed in the accreting 2.5 ms pulsar SAX J1808.4-3658
\cite{rudy}. Other excitations, notably when the difference of epicyclic
frequencies is equal to the pulsar spin frequency, are also possible.
This would explain why in many neutron-star kHz QPO sources, the
two QPO frequencies are separated either by one or one-half
the inferred spin frequency \cite{vdk}.

In conclusion, we concur that quasi-periodic modulations of the
X-ray flux in low-mass X-ray binaries reflect oscillations
of the accretion disk \cite{wagoner,kato}. However, non-linear effects
must be taken into account.

%%%%%%%%%%%%%%%%%%%%%%%%%%%%%%%%%%%%%%%%%%%%%%%%
%% End
%%%%%%%%%%%%%%%%%%%%%%%%%%%%%%%%%%%%%%%%%%%%%%%%

\begin{theacknowledgments}
Research supported in part by the Polish KBN with grant 2P03D01424,
and the European Commission grant {\it Access to Research Infrastructure
 action of the Improving Human Potential Program} to the UK
Astrophysical Fluids Facility in Leicester University, where this contribution
has been completed. With thanks to
V.~Karas,
S.~Kato,
M.~van~der~Klis,
J.-P.~Lasota,
C.~Mauche,
P.~Rebusco,
N.~Stergioulas, 
and
R.~Wagoner,
for their illuminating suggestions.
We thank Gabriel T\"or\"ok for preparing Fig.~1.
\end{theacknowledgments}

\end{document}